*Perspective*

# Printing via Laser-Induced Forward Transfer and the Future of Digital Manufacturing

Camilo Florian [1,2,*]  and Pere Serra [3]

1. Princeton Institute for the Research and Technology of Materials (PRISM), Princeton University, 70 Prospect Av, Princeton, NJ 08540, USA
2. Instituto de Óptica Daza de Valdés, Consejo Superior de Investigaciones Científicas (IO-CSIC), Calle Serrano 122, 28006 Madrid, Spain
3. Departament de Fisica Aplicada, Universitat de Barcelona, Martí i Franqués 1, 08028 Barcelona, Spain
* Correspondence: camilo.florian@csic.es

**Abstract:** In the last decades, digital manufacturing has constituted the headline of what is starting to be known as the 'fourth industrial revolution', where the fabrication processes comprise a hybrid of technologies that blur the lines between fundamental sciences, engineering, and even medicine as never seen before. One of the reasons why this mixture is inevitable has to do with the fact that we live in an era that incorporates technology in every single aspect of our daily lives. In the industry, this has translated into fabrication versatility, as follows: design changes on a final product are just one click away, fabrication chains have evolved towards continuous roll-to roll processes, and, most importantly, the overall costs and fabrication speeds are matching and overcoming most of the traditional fabrication methods. Laser-induced forward transfer (LIFT) stands out as a versatile set of fabrication techniques, being the closest approach to an all-in-one additive manufacturing method compatible with virtually any material. In this technique, laser radiation is used to propel the material of interest and deposit it at user-defined locations with high spatial resolution. By selecting the proper laser parameters and considering the interaction of the laser light with the material, it is possible to transfer this technique from robust inorganic materials to fragile biological samples. In this work, we first present a brief introduction on the current developments of the LIFT technique by surveying recent scientific review publications. Then, we provide a general research overview by making an account of the publication and citation numbers of scientific papers on the LIFT technique considering the last three decades. At the same time, we highlight the geographical distribution and main research institutions that contribute to this scientific output. Finally, we present the patent status and commercial forecasts to outline future trends for LIFT in different scientific fields.

## 1. Introduction

Digital manufacturing comprises fabrication approaches and techniques to create functional structures and devices where the design, modification, and final optimization can be performed digitally. The biggest advantage over conventional fabrication is the possibility to work without masks, master samples, or molds, allowing design correction in parallel with fabrication. In general, these techniques constantly try to solve key open challenges, such as low overall fabrication costs, industrial scalability, and compatibility with current techniques, wide material compatibility, and high spatial resolution and reproducibility. Researchers and engineers pursue the creation of an all-in-one technique that tackles these challenges but, unfortunately, this is an elusive task, since in most of the cases, there are limitations in at least one of these fronts. For example, despite the fact that there are numerous fabrication techniques with good performances in laboratory conditions, one of the biggest problems still comes when these fabrication techniques are required to perform in real industrial workspaces, where scalability and mass production at the lowest costs and highest speeds mark clear quests for developers.

In the present paper, we center our attention on a specific laser printing technique commonly known as laser-induced forward transfer (LIFT). In laser printing via LIFT, irradiation from a laser source is focused on a donor substrate that contains the material of interest. The interaction of the radiation with the donor material results in the propulsion of a tiny amount of material towards an acceptor substrate conveniently positioned, giving a place for the formation of a printed voxel. Through the sequential printing of voxels, it is possible to generate practically any layout. Originally, the technique was mostly suitable for pulsed laser sources using mainly solids as donor material, which limited its scope but, nowadays, the increasingly wider availability of reliable and affordable laser sources (either pulsed or CW, with different wavelengths), combined with the adequate material preparation (solid, paste, or liquid form), it is possible to create devices for the microelectronics industry, as well as miniaturized tests for microbiology, and it has lately shown potential for tissue engineering applications, among many others. Importantly, this paper does not intend to be the ultimate review of the technique (for that, we will direct the reader to corresponding review papers), and neither is it an open discussion on the most recent advances. We start the introduction by surveying review papers on the current developments and sky edge applications that implement this printing technique, such as the ones in references [1–3]. From there, the principle of operation for the transfer of materials disposed in solid, paste, and liquid films for diverse applications is discussed. With this information in hand, we then provide a general assessment on the past, present, and future of the LIFT technique considering published patents and patent applications where the LIFT is the central technique. Finally, by analyzing commercial data from the broader picture of additive manufacturing, we present commercial forecasts and possible trends for the LIFT technique in exploitable scientific areas, such as electronics, biology, and medicine, to name a few.

## 2. A Brief Overview of the Laser-Induced Forward Transfer (LIFT) Technique

In the printing of materials via laser-induced forward transfer (LIFT), the material of interest is generally disposed as a thin film onto a transparent substrate called the "donor substrate". A laser beam is then focused at the interface between the thin film and the holding transparent substrate, producing a micro-explosion that propels the material forward. When another "receptor substrate" is positioned at a convenient distance from the donor, the propelled material is safely deposited at specific user-defined positions. Traditionally, a pulsed laser is used, and each pulse produces a printed voxel of material, the volume of which volume can be as low as a few femtoliters [4]. Recent developments have implemented approaches that allow the printing of materials with two or more laser pulses [5,6], and even continuous wave (CW) laser sources [7]. The resulting pixel geometry and amount of transferred material depend on parameters regarding the initial conditions of the donor material (thickness, viscosity, optical properties), the use of an additional sacrificial absorbing layer (commonly known as a dynamic release layer or 'DRL'), the physical and chemical properties of the receptor substrate (wetting, oxidation, temperature), and the laser conditions (fluence, intensity distribution, wavelength, pulse duration). Figure 1 contains a sketch that displays the general configuration and main elements in a traditional configuration to transfer materials via LIFT.

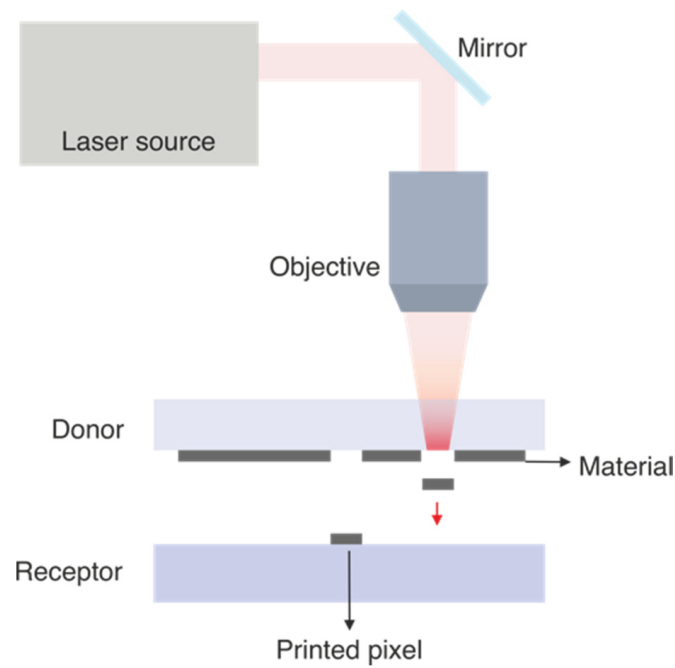

**Figure 1.** Sketch (not to scale) that includes the main elements of an experimental setup to print materials via LIFT.

In order to provide a classification scheme for all the LIFT variants, we can use as a sorting parameter the disposition of the material to be transferred, being (i) materials in a solid state, (ii) pastes, (iii) or simply liquids. In the following, we will briefly discuss the principles of operation for each case and provide relevant examples for each case.

*2.1. LIFT from Solids*

In this group of applications for the LIFT technique, the starting base material in the donor film corresponds to a solid layer, that usually absorbs the incoming radiation and is expelled towards the receptor substrate [1]. Two general scenarios could be drawn depending on how the material is modified by the laser radiation; in the first one, the solid material is locally melted, and it is the liquid phase that is propelled forward [8]. The solidification occurs generally on the receptor substrate in the form of isolated hemispherical droplets, as displayed in Figure 2A. A second case takes place when the material to be transferred should not undergo any material phase change. This particular approach is suitable for extremely sensitive materials where the crystalline structures should be conserved after the transfer, or the material is highly sensitive to thermal changes induced by the laser irradiation. In this case, the use of an intermediate sacrificial layer is implemented. This layer will absorb most of the incoming radiation, producing the localized explosion that will propel the material of interest towards the receptor substrate [9]. The transfer dynamics in this case is more complex, since it involves the participation of an additional material that will react differently to the laser radiation, as displayed in Figure 2B.

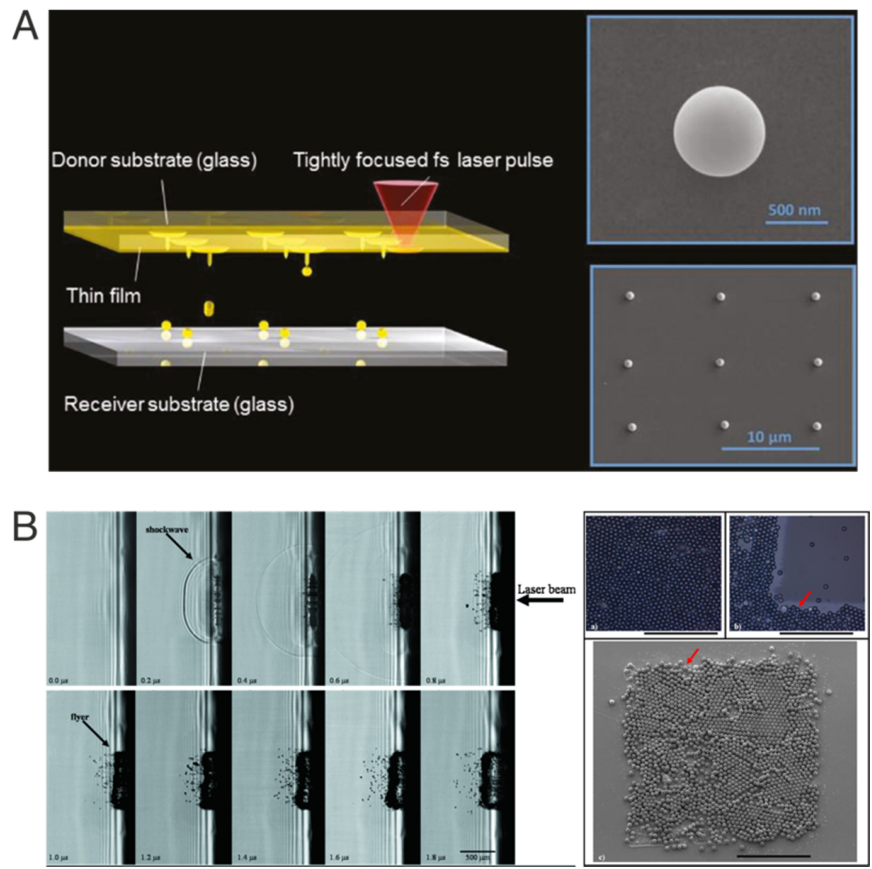

**Figure 2.** Printing of materials starting from a solid donor film following two different scenarios. (**A**) Melted material, where the final output corresponds to individual spherical particles. Reprinted under creative commons permission (CC BY-NC-ND 4.0) from [8], copyright 2018, De Gruyter. (**B**) In this transfer, the use of a sacrificial layer allowed the printing of an array of microspheres that followed the square shape distribution of the incident laser beam. Reprinted with permission from [9], with the permission of AIP Publishing (Copyright 2010).

*2.2. LIFT from Pastes*

In this particular case of the LIFT technique, also known as laser decal transfer (LDT), the donor material is a highly viscous paste. Due to the nature of both material and processing conditions, the paste is not subjected to deformation due to surface tension effects. Typically, the transferred material shape has the same geometry as the laser intensity distribution, which allows the transfer of voxels of material with complex shapes when a regular LIFT setup is combined with digital micromirrors, spatial light modulators, or phase masks to define a particular intensity distribution profile. In the example shown in Figure 3 [10], the donor material corresponds to a silver nanoparticle suspension with a viscosity of 100 Pa.s. Congruent transfers occur when the right viscosity is achieved. The laser radiation impinges the donor film from the top and the interaction is produced at the interface between the transparent holder substrate and the paste film. In the figure, each image corresponds to a frame of a video acquired with a high-speed camera at the times indicated in the insets. The whole transfer process corresponds to a single event. When a receiving substrate is placed close by, the resulting deposition corresponds to a flat cylinder in the form of a wafer, the thickness of which is similar to the one of the donor films.

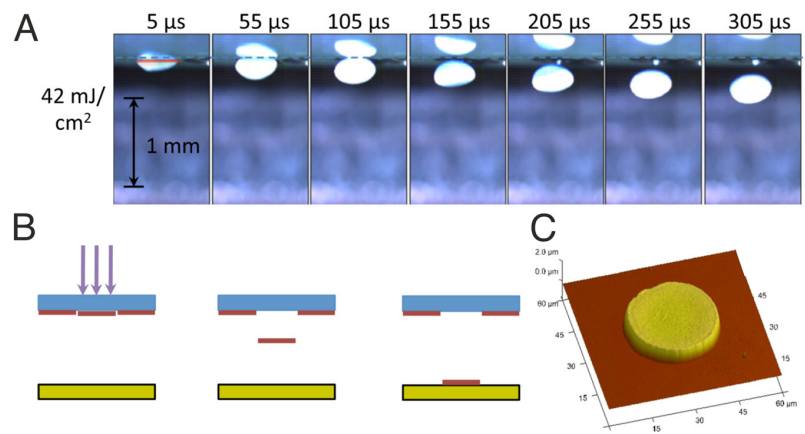

**Figure 3.** (**A**) Snapshots of a high-speed video of the transfer via LIFT of a high viscosity silver paste. In the top figure, a single laser pulse with a circular shape comes from the top, pushing downwards a 'coin-like' voxel that flies towards the receptor substrate without breaking. The schematics is included in (**B**), and a topography acquired via AFM of the final printed material is shown in (**C**). The figure has been reprinted from [10], with the permission of AIP Publishing (Copyright 2013).

*2.3. LIFT from Liquids*

A plethora of variants exist when the material to be printed is disposed as a liquid film in the donor substrate [1,3]. The transfer dynamics are similar to the previous examples; however, there are additional parameters to consider, particularly regarding the energy absorption in the liquid and the rheological conditions of the material. In general, once the laser is absorbed by the liquid or the intermediate absorbing layer, a bubble is produced, followed by the formation of a jet that travels at high speed towards the receptor substrate conveniently placed in the vicinity of the donor substrate. The liquid jet gently feeds what will finally turn into a printed pixel in the shape of a hemispherical droplet, the dimensions of which depend on the amount of transferred material and the surface energy of the liquid on the surface of the receptor substrate. Figure 4 shows examples of LIFT of liquids with no use of an intermediate layer (Figure 4A [11]), with a metallic absorbing layer (Figure 4B [12]), and with the use of an absorbing polymer layer (Figure 4C [13]). More detailed examples can be found in the most recent review articles on the technique [1,3].

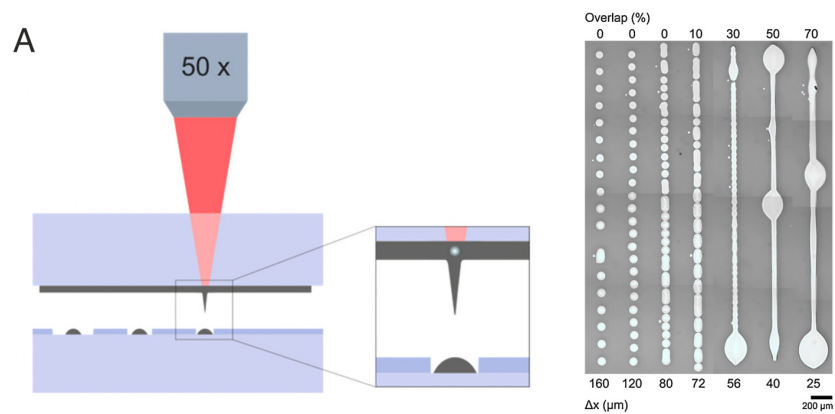

**Figure 4.** *Cont.*

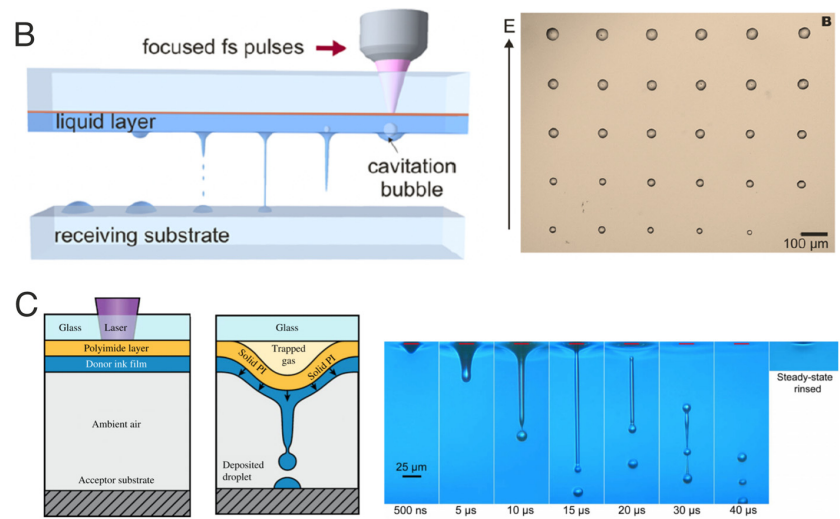

**Figure 4.** (**A**) No sacrificial layer (the liquid absorbs the laser radiation). Reprinted from [11], copyright (2016), with permission from Elsevier. (**B**) Metallic intermediate layer. Reprinted from [12], copyright (2017), with permission from Elsevier. (**C**) Blister-actuated LIFT based on the use of a polymer absorbing layer. Reprinted/adapted from Ref. [13] (Copyright 2018) with permission from Wiley Books.

## 3. LIFT Scientific Literature

The first report on the idea of using a laser beam to propel materials towards a receptor substrate was published by Levene et al. in 1970 [14]. In that experiment, the donor substrate was a polyethylene typewriter ribbon covered with black ink. Material deposition was reported to be feasible for both forward (such as in Figure 1) and backward configuration (originally called 'reverse transfer', where the material is ejected in the opposite direction to the laser beam), producing continuous printed lines of ink. It was not until 1986 that Bohandy et al. [15] coined the term LIFT (laser-induced forward transfer) in what became one of the most cited papers on the field. Since then, numerous reports have been published, accounting for 661 peer-reviewed papers, as can be seen in the plot of Figure 5, with an increasing number of citations according to the ISI Web of Science database.

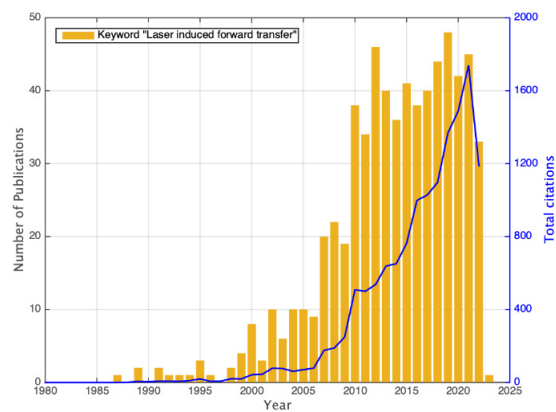

**Figure 5.** Number of published peer-reviewed articles (grand total of 661) and total number of citations per year. Data obtained via the ISI Web of Science, by sorting the keyword "laser induced forward transfer" as of 11 November 2022.

It has to be noted that, in most cases, when talking about transferring materials via LIFT, the material to be deposited is virtually the same as that disposed in the donor substrate. In the cases where the material is significantly altered compared to the donor material, we have only included studies called reactive LIFT (r-LIFT), in which the material changes its chemical composition mainly due to the action of traveling through a reactive atmosphere or by producing certain structural change due to the laser radiation [16]. It is not the case for the papers regarding matrix-assisted pulsed laser evaporation (MAPLE), which also follows the printing principle of LIFT, but that are not included in the data of Figure 5. The MAPLE technique has grown significantly and independently, generating a full string of numerous papers that deserve their own study; for this reason, they are out of scope in the present paper.

Among the publications included in Figure 5, the top five journals where the research was published were Applied Surface Science (13%), Proceedings of SPIE (9%), Applied Physics A (7%), Applied Physics Letters (3.6%), and the Journal of Laser Micro Nanoengineering (3.2%). An extended list is included in Table 1. Using the classification proposed by the ISI Web of Science, the fields which these publications fall into include applied physics, materials science, optics, physical chemistry, and condensed matter.

**Table 1.** Journal names and number of papers published on LIFT, according to the ISI Web of Science, by sorting the keyword "laser induced forward transfer" as of 11 November 2022.

| Publication Name | Number of Papers | Percentage (%) |
| --- | --- | --- |
| Applied Surface Science | 84 | 12.7 |
| Proceedings of the Society of Photo Optical Instrumentation Engineers SPIE | 59 | 8.9 |
| Applied Physics A Materials Science Processing | 48 | 7.2 |
| Applied Physics Letters | 24 | 3.6 |
| Journal of Laser Micro Nanoengineering | 21 | 3.1 |
| Optics Express | 14 | 2.1 |
| Journal of Applied Physics | 11 | 1.6 |
| Conference on Lasers and Electro Optics | 9 | 1.3 |
| Journal of Physics D Applied Physics | 9 | 1.3 |
| Nanomaterials | 9 | 1.3 |

We have plotted the same data of Figure 5 in a map displayed in Figure 6. The number of publications is sorted by the country where the corresponding author institution was reported. From here, we can see how the most productive region overall is Europe, with notable contributions from North America and Asia. An interactive tool is included as a Supplementary Material.

Table 2 contains a list with the names of the main funding agencies that made possible most of the contributions of the researchers on the field in these countries, including the number of papers produced and its percentage with respect to the total number of publications (661). This information is also available at the ISI Web of Science database.

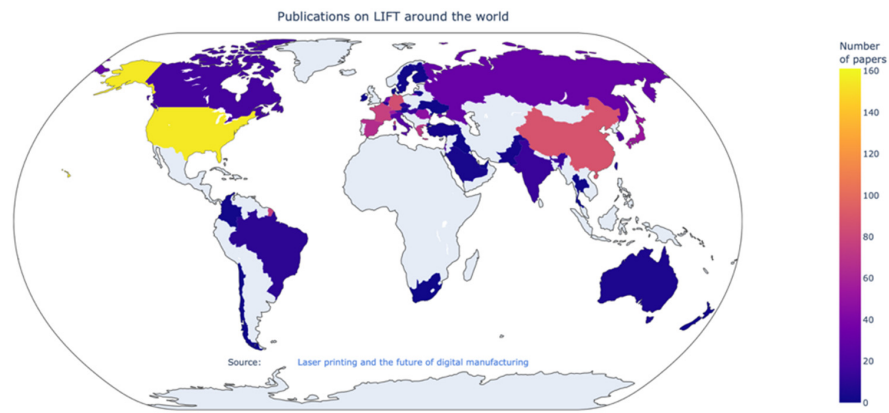

**Figure 6.** Geographical distribution of the total amount of publications per country on LIFT around the world. Data obtained via the ISI Web of Science by sorting the keyword "laser induced forward transfer" as of 8 June 2022. An interactive resource is available as a Supplementary Material.

**Table 2.** Main international funding agencies that support the research reflected in the published papers on LIFT, according to the ISI Web of Science, by sorting the keyword "laser induced forward transfer" as of 11 November 2022.

| Institution | Country | Number of Papers | Percentage (%) |
|---|---|---|---|
| European Commission | European countries | 99 | 14.9 |
| National Natural Science Foundation of China | China | 34 | 5.1 |
| National Science Foundation | United States | 25 | 3.7 |
| German Research Foundation | Germany | 24 | 3.6 |
| Office of Naval Research | United States | 23 | 3.4 |
| Engineering Physical Sciences Research Council | United Kingdom | 21 | 3.1 |
| UK Research Innovation | United Kingdom | 21 | 3.1 |
| French National Research Agency | France | 19 | 2.8 |
| Spanish Government | Spain | 19 | 2.8 |
| Swiss National Science Foundation | Switzerland | 13 | 1.9 |

## 4. Industrial Perspectives

### 4.1. Patent Applications Landscape

Relevant technological throughputs are often published as patents in order to grant intellectual property rights to the authors. It is important to note that patent applications and the process for their publication follows a procedure that is dramatically different from the one of scientific publications, and technical language is often used in combination with economic and legal terms, in order to provide a solid legal ground for protecting the technology as much as possible. For this reason, we focus our attention on patents that contain the keywords "laser induced forward transfer" from two public patent office databases, considering that the majority of publications on LIFT are located in these two regions according to Figure 6. These patent office databases are the European Patent Office (EPO) and the United States Patent and Trademark Office (USPTO). Based on this premise, we select patents publications that are publicly accessible and published with an assigned patent publication number, regardless of the acceptance or licensing status. In the following data we, therefore, jointly present patents with patent applications. We find 321 patents reported in the EPO and 303 in the USPTO, that, regardless of the publication country, are reported in these databases. Due to the large number of items, the specific lists are included as a Supplementary Material. The following plot in Figure 7 shows the number of patents and patent publications over time, considering the early publication date for the EPO results and the publication date for the USPTO results.

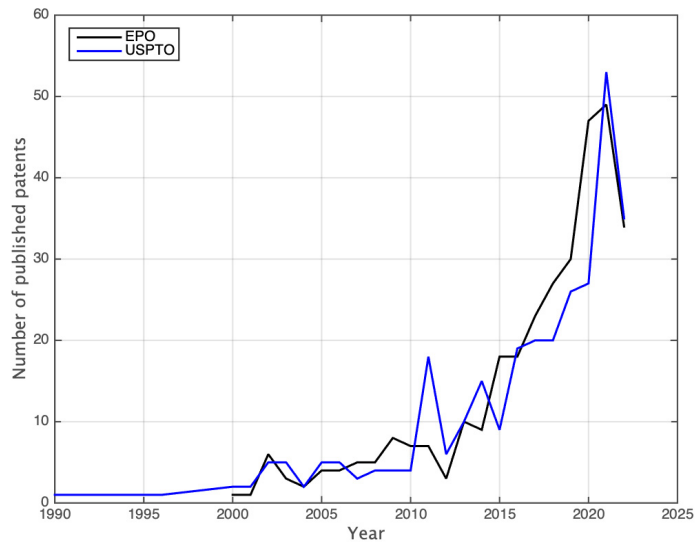

**Figure 7.** Number of published patent applications reported in the European Patent Office (EPO) and the United States Patent and Trademark Office (USPTO) per year. Data obtained on the respective databases, by sorting the keyword "laser induced forward transfer" as a whole string on 11 November 2022.

*4.2. Commercial Forecast and Future Trends*

From the current report on "The worldwide market for lasers: market review and forecast 2020" [17] from the European Photonics Industry Consortium (EPIC), that includes projections from 2020 to 2025 on different industries where laser processing of materials is at the center, we can extract an overview for future projections from the economical point of view.

We identify three main areas of revenue where laser printing can find a niche for development into industrial applications, including all the available commercial applications. As a disclosure note, these are just a limited selection of the biggest commercial areas. These areas are additive manufacturing, laser-based paper printing, and medicine. In the case of additive manufacturing, the reported total gross revenue for all applications accounted for 75.3M US dollars in 2022. It is projected a growth of 13.1% over the current decade. The second area of laser-based paper printing accounted for a revenue of around 48.5M US dollars for the year 2022; however, in this case, it is forecast that laser-based technology revenue in this commercial area will decrease over 4% over the current decade, mainly due to a shift toward a paper-less work culture that will mean a decrease in sales of printers in general. Finally, for the area of medicine, the revenue accounted for about 1500M US dollar in 2022, with an average growth estimation of about 7.4% over the current decade. For the LIFT technique, these figures are encouraging, particularly since the characteristic properties of the technique are suitable for the printing of complex materials in all the mentioned areas, as we will discuss in the following.

In the area of additive manufacturing, the selective deposition of materials to fabricate three-dimensional structures has been commercially exploited thanks to important developments in techniques and machines for the printing of polymers, metals, and ceramic-based materials. Commercial 3D printers are one successful example that has been booming in the last decade, and allows us to produce easily accessible methods for functional devices without the need for expensive lab equipment. In these printers, the material of interest is generally deposited in a layer-by-layer fashion, and it is subsequently melted at user-defined locations due to the interaction with an energetic laser beam [18–20]. At the end, several two-dimensional stacked layers lead to the fabrication of three-dimensional solid structures. The fundamental operation of these 3D printers has been applied for the additive deposition of diverse materials, allowing the fabrication of glass and polymer lab-on-a-chip devices with micrometer resolutions [21,22], as well as the construction of resin boats [23] and metallic rocket fuel tanks [24] of exceptional quality. The same principles for printing are used with the LIFT technique. There are numerous publications where the printing of solids, pastes, and liquid containing conductive materials have been used for the fabrication of conductive paths [25–27], light-emitting diodes [28], solar cells [29], and materials used for microsensors [30]. An important advantage over other printing techniques, such as ink-jet printing, relies on the fact that these systems are able to work with a wider viscosity range of inks with nanoparticle suspensions, which opens a wider variety of printable materials suitable for electronic applications, especially in the case of high-viscosity pastes made from particles with particle sizes in the microscale [1].

In the area of laser-based paper printing, there is one example of a commercial initiative that used the LIFT technique for graphical design paper printing on large areas. The machine was named Lasersonic, and it was capable of printing inks on paper at rates of 1.3 $m^2$/min, with resolutions of 600 dpi [31]. In this application, the ink used was disposed as a thin liquid film, the thickness of which was regulated by cylinders that provided a virtually unlimited ink supply, allowing for printing the ink at such high speeds in a roll-to-roll setup, demonstrating compatibility with current industrial fabrication techniques. The same principle can be applied for the commercial printing of flexible substrates, largely used in electronic devices, such as sensors [32], displays [33], electronic components [34], and solar cells [35], to name a few. In addition, it has been recently demonstrated that the printing of conductive inks via LIFT on paper can also be used for the fabrication of low-cost electronic devices [32].

In the area of medicine, the LIFT technique is of particular interest in the development of tissue printing for the fabrication of organs and organelles [36–40]. A success commercial example is the company Poietis, whose 4D Bioprinting technique incorporates a LIFT device into a commercial bioprinting machine for single-cell transfer. The advantages of the LIFT technique regarding positioning and controlled volume deposition, in combination with an in situ optical microscope, allows this machine to track over time the growth of individual cells and extracellular matrices. The company founders have different publications on the characterization of the technique towards its usage in bioprinting [41], as well as different patents where the LIFT technique is the core approach for the transfer of different biological materials. Although there are open challenges to be overcome, such as open questions regarding donor film drying (in the case of liquid donor films), cell aggregation, and potential metal toxicity, among others [38], there are advances on the field where the LIFT technique has been shown as useful for the transfer of biological materials [36–39]. It is also interesting to point out that by sorting out the strings "laser induced forward transfer" with "organ" and "organelle" in the EPO patent database, just in the last decade, 45 + 12 patents have been filed, which constitutes 17.7% of all of the patents data shown in Figure 7 for the EPO results. This demonstrates the exploitation potential for future technologies and a crescent interest not only from researchers, but also from companies in the medical and biological areas.

**Supplementary Materials:** The following supporting information can be downloaded at: https://www.mdpi.com/article/10.3390/ma16020698/s1.

**Author Contributions:** C.F. wrote the original draft. C.F. and P.S. edited the manuscript. All authors contributed to the scientific discussion and revision of the article. All authors have read and agreed to the published version of the manuscript.

**Funding:** C.F. acknowledges the support from the European Commission through the Marie Curie Individual Fellowship—Global grant No. 844977, while P.S. acknowledges the support from project PID2020-112669GB-I00 funded by MCIN/AEI/10.13039/501100011033.

**Institutional Review Board Statement:** Not applicable.

**Informed Consent Statement:** Not applicable.

**Data Availability Statement:** The data presented in this paper is available upon request to the corresponding author.

**Acknowledgments:** Special thanks to Antonio Castello from EPIC for helpful discussions.

**Conflicts of Interest:** The authors declare no conflict of interest.